\documentclass[12pt]{iopart}
\usepackage{epsfig}
\usepackage[T1]{fontenc}
\usepackage{ae,aecompl}
\usepackage{iopams}

\begin{document}

\title{Material-Point Simulation to Cavity Collapse Under Shock}
\author{Aiguo Xu, X. F. Pan, Guangcai Zhang and Jianshi Zhu}
\address{Laboratory of Computational Physics, \\
Institute of Applied Physics and Computational Mathematics, P. O.
Box 8009-26, Beijing 100088, P.R.China} \ead{Xu\_Aiguo@iapcm.ac.cn}
\begin{abstract}
The collapse of cavities under shock is a key problem in various
fields ranging from erosion of material, ignition of explosive, to
sonoluminescence, etc. We study such processes using the
material-point-method developed recently in the field of solid
physics. The main points of the research include the relations
between symmetry of collapsing and the strength of shock, other
coexisting interfaces, as well as hydrodynamic and thermal-dynamic
behaviors ignored by the pure fluid models. In the case with strong
shock, we study the procedure of jet creation in the cavity; in the
case with weak shock, we found that the cavity can not be collapsed
completely by the shock and the cavity may collapse in a nearly
isotropic way. The history of collapsing significantly influences
the distribution of ``hot spots" in the shocked material. The change
in symmetry of collapsing is investigated. Since we use the Mie-Gr%
\"{u}neisen equation of state and the effects of strain rate are not
taken into account, the behavior is the same if one magnifies the
spatial and temporal scales in the same way.
\end{abstract}


\vspace{2pc}
\submitto{\JPCM}\maketitle

\section{Introduction}

 The first work on cavitation phenomena dated
back to 1894 when Reynolds observed such a phenomenon in water
flowing through a tube with a local constriction\cite{1894}. A few
years later, very destructive actions of cavitation to some
industrial systems, such as steamship propellers, hydroturbine, etc,
were found\cite{1912,1944,1,B2002}. Then, it was addressed and
discussed under other various backgrounds\cite{5,6,7,8}, for
example, hot spots in explosives from which reaction
ensues\cite{B2002}, the recent observation of
sonoluminescence\cite{Putterman1995}, etc. Now, controlled
cavitation erosion has been regarded as a powerful tool for modern
technologies like steam/wet laser cleaning\cite{2}, effective
salmonella destruction\cite{3}, and treatment for kidney
stones\cite{4}. Among the publications on cavitation in literature,
most of them resorts to experiments, then theoretical calculations,
and only a small portion rely on numerical simulations. The physical
models used can be put into two categories, fluidic ones and solid
ones. The numerical tools for the former are hydrodynamic
codes\cite{HYDRO}, and for the latter are molecular
dynamics(MD)\cite{ZYH2006,YQL2007,MD} and finite element
method\cite{FE}.

Most of the analysis here are based on fluidic models. The most elegant
mathematical description of bubble motion is one-dimensional\cite{B2002}. It
has long been observed that a tongue of liquid may be projected into a
bubble under gravity and a tongue of metal may be projected into a cavity
under shock, which confirm that one primary mechanism of erosion by
cavitation is mechanical\cite{1944,R1917}. The observation of such an
asymmetric collapse stimulates the question as to what kind of boundary
conditions give rise to it. The first kind of attempt simply considers it to
be a geometric flow effect of the boundary. The second one regards it as the
spallation of one surface towards the other by an incident shock, where the
upstream face converges towards the stationary downstream one in the jet. In
either of the two cases, the aspherical collapse implies that an impulse is
applied, typically by an impact jet, to the surrounding fluid or solid in
contact\cite{B2002}.

From the numerical simulation side, fluidic models on cavitation in metals
ignore some important characteristics of solid, such as plastic strain,
hardening, and properties relevant to the deformation history. At the same
time, the material under investigation is highly distorted during the
collapsing of cavities. This causes severe problems in computational
modeling of such a process. It is well known that the Eulerian description
is not convenient to tracking interfaces. When the Lagrangian formulation is
used to describe problems with large displacement and large strain, the
original element mesh becomes distorted so significantly that the mesh has
to be re-zoned to restore proper shapes of elements. The state fields of
mass density, velocities and stresses must be mapped from the distorted mesh
to the newly generated one. This mapping procedure is not a straightforward
task, and introduces errors. To treat with large distortions problems,
several mixed methods have been developed recently. Among these methods, the
material-point method (MPM), introduced originally in fluid dynamics by
Harlow, et al\cite{H1964} and extended to solid mechanics by Burgess, et al
\cite{MPM}, overcomes the above-mentioned troubles. At each time step,
calculations consist of two parts: a Lagrangian part and a convective one.
Firstly, the computational mesh deforms with the body, and is used to
determine the strain increment, and the stresses in the sequel. Then, the
new position of the computational mesh is chosen (particularly, it may be
the previous one), and the velocity field is mapped from the particles to
the mesh nodes. Nodal velocities are determined using the equivalence of
momentum calculated for the particles and for the computational grid. The
method not only takes advantages of both the Lagrangian and Eulerian
algorithms but makes it possible to avoid their drawbacks as well. In this
work we use a improved MPM\cite{PXFGF2007} to investigate the collapse of
cavities in shocked metal.

The following part of the paper is organized as follows. Section 2
describes briefly the numerical scheme. Section 3 shows the physical
model and section 4 presents simulation results. Section 5 makes the
conclusion.

\section{The Material-Point Method}

The material-point method\cite{MPM} is a particle method, where the
continuum bodies are discretized with $N_{p}$ material particles.
Each material particle
carries the information of position $\mathbf{x}_{p}$, velocity $\mathbf{v}%
_{p}$, mass $m_{p}$, density $\rho _{p}$, stress tensor
$\boldsymbol{\sigma}_p$ , strain tensor $\boldsymbol{\varepsilon
}_{p}$ and all other internal state variables necessary for the
constitutive model, where $p$ is the index of particle. At each time
step, the mass and velocities of the material particles are mapped
onto the background computational mesh. The mapped
momentum at node $i$ is obtained by $m_{i}\mathbf{v}_{i}=\sum_{p}m_{p}%
\mathbf{v}_{p}N_{i}(\mathbf{x}_{p})$, where $N_{i}$ is the element shape
function and the nodal mass $m_{i}$ reads $m_{i}=\sum_{p}m_{p}N_{i}(\mathbf{x%
}_{p}).$ Suppose that a computational mesh is constructed of eight-node
cells for three-dimensional problems, then the shape function is defined as%
\begin{equation}
N_{i}=\frac{1}{8}(1+\xi \xi _{i})(1+\eta \eta _{i})(1+\varsigma
\varsigma _{i})\texttt{,}  \label{MPMe1}
\end{equation}%
where $\xi $,$\eta $,$\varsigma $ are the natural coordinates of the
material particle in the cell along the x-, y-, and z-directions,
respectively, $\xi _{i}$,$\eta _{i}$,$\varsigma _{i}$ take corresponding
nodal values $\pm 1$. The mass of each particle is equal and fixed, so the
mass conservation equation, $\mathrm{d}\rho /\mathrm{d}t+\rho \nabla \cdot
\mathbf{v}=0$, is automatically satisfied.

For pure mechanical problems the differential equation of balance
reads,
\begin{equation}
\rho
\mathrm{d}\mathbf{v/}\mathrm{d}t=\nabla \cdot \boldsymbol{\sigma }+\rho \mathbf{b%
}\texttt{,} \label{MPMeq2}
\end{equation}
 where $\rho $ is the mass density, $\mathbf{v}$ the velocity, $%
\boldsymbol{\sigma }$ the stress tensor and $\mathbf{b}$ the body
force. The momentum equation is solved on a finite element mesh in a
lagrangian frame. The weak form of it can be found, based on the
standard procedure,
\begin{equation}
\begin{array}{ll}
 & \int_{\Omega }{ \rho \delta \mathbf{v}\cdot \mathrm{d}\mathbf{v/}\mathrm{d}t%
\mathrm{d}\Omega }+ \int_{\Omega }{\delta
(\mathbf{v}\nabla) \cdot \boldsymbol{\sigma}\mathrm{d}\Omega } - \int_{\Gamma _{t}}{\ \delta \mathbf{v}\cdot \mathbf{t}\mathrm{d}\Gamma }\\
 & -\int_{\Omega
}{\ \rho \delta \mathbf{v}\cdot \mathbf{b}\mathrm{d}\Omega } =0
\texttt{.} \label{1}
\end{array}
\end{equation}
Since the continuum bodies is described with the use of a finite set
of material particles, the mass density can be written as $\rho
(\mathbf{x})=\sum_{p=1}^{N_{p}}{\ m_{p}\delta
(\mathbf{x}-\mathbf{x}_{p})}$, where $\delta $ is the Dirac
delta function with dimension of the inverse of volume. The substitution of $%
\rho (\mathbf{x})$ into the weak form of the momentum equation converts the
integral to the sums of quantities evaluated at the material particles,
namely,
\begin{equation}
m_{i}\mathrm{d}\mathbf{v}_{i}/\mathrm{d}t=(\mathbf{f}_{i})^{\mathrm{int}}+(%
\mathbf{f}_{i})^{\mathrm{ext}}\texttt{,}  \label{MPMe2}
\end{equation}
where the internal force vector is given by $\mathbf{f}_{i}{}^{\mathrm{int}%
}=-\sum_{p}^{N_{p}}{m_{p}\boldsymbol{\sigma }}_{p}{\cdot (\nabla N_{i})/\rho _{p}%
}$, and the external force vector reads $\mathbf{f}_{i}{}^{\mathrm{ext}%
}=\sum_{p=1}^{N_{p}}{N_{i}\mathbf{b}_{p}+\mathbf{f}_{i}^{c}}$, where the
vector $\mathbf{f}_{i}^{c}$ is the contacting force between two bodies. In
present paper, all colliding bodies are composed of the same material, and $%
\mathbf{f}_{i}^{c}$ is treated with in the same way as the internal
force.

The nodal accelerations are calculated by Eq. (\ref{MPMe2}) with an
explicit time integrator. The critical time step satisfying the
stability conditions is the ratio of the smallest cell size to the
wave speed. Once the motion equations are solved on the cell nodes,
the new nodal values of acceleration are used to update the velocity
of the material particles. The strain increment for each material
particle is determined using the gradient of nodal basis function
evaluated at the position of the material particle. The
corresponding stress increment can be found from the constitutive
model. The internal state variables can also be completely updated.
The computational mesh may be the original one or a newly defined
one, choose for convenience, for the next time step. More details of
the algorithm is referred to \cite{PXFGF2007,Ma}.


\section{Physical model}

We consider an associative von Mises plasticity model with linear
kinematic and isotropic hardening\cite{CModel}. Introducing a linear
isotropic elastic relation, the volumetric plastic strain is zero,
leading to a deviatoric-volumetric
decoupling. So, it is convenient to split the stress and strain tensors, $%
\boldsymbol{\sigma }$ and $\boldsymbol{\varepsilon }$, as
\begin{eqnarray}
\boldsymbol{\sigma } &=&\mathbf{s}-P\mathbf{I},P=-\frac{1}{3}\verb"Tr"(\boldsymbol{\sigma }%
)\texttt{,}  \label{PMe1} \\
\boldsymbol{\varepsilon } &=&\mathbf{e}+\frac{1}{3}\theta \mathbf{I},\theta =%
\frac{1}{3}\verb"Tr"(\boldsymbol{\varepsilon })\texttt{,}
\label{PMe2}
\end{eqnarray}%
where $P$ is the pressure scalar, $\mathbf{s}$ the deviatoric stress tensor,
and $\mathbf{e}$ the deviatoric strain. The strain $\mathbf{e}$ is generally
decomposed as $\mathbf{e}=\mathbf{e}^{e}+\mathbf{e}^{p}$, where $\mathbf{e}%
^{e}$ and $\mathbf{e}^{p}$ are the traceless elastic and plastic components,
respectively. The material shows a linear elastic response until the von
Mises yield criterion,
\begin{equation}
\sqrt{\frac{3}{2}}\left\Vert \mathbf{s}\right\Vert =\sigma
_{Y}\texttt{,} \label{PM1}
\end{equation}%
is reached, where $\sigma _{Y}$ is the plastic yield stress. The yield $%
\sigma _{Y}$ increases linearly with the second invariant of the plastic
strain tensor $\mathbf{e}^{p}$, i.e.,
\begin{equation}
\sigma _{Y}=\sigma _{Y0}+E_{\tan }\left\Vert
\mathbf{e}^{p}\right\Vert \texttt{,} \label{PM4}
\end{equation}%
where $\sigma_{Y0}$ is the initial yield stress and $E_{\tan }$ the
tangential module. The deviatoric stress $\mathbf{s}$ is calculated by
\begin{equation}
\mathbf{s}=\frac{E}{1+\nu }\mathbf{e}^{e} \texttt{,} \label{s}
\end{equation}
where $E$ is the Yang's module and $\nu$ the Poisson's ratio. Denote
the initial material density and sound speed by $\rho_0$ and $c_0$,
respectively. The shock speed $U_s$ and the particle speed $U_p$
after the shock follows a linear relation, $U_s = c_0 + \lambda
U_p$, where $\lambda$ is a characteristic coefficient of material.
The pressure $P$ follows the Mie-Gr\"{u}neisen equation of state
\begin{equation}
P=(C\mu +D\mu ^{2}+S\mu ^{3})(1-\frac{\gamma \mu }{2})+\gamma \rho
\epsilon \texttt{,} \label{Peos}
\end{equation}
where $\epsilon $ is the internal energy, $\mu =\rho /\rho _{0}-1$, $\gamma
=\gamma _{0}\rho _{0}/\rho $, $C=\rho _{0}c_{0}^{2}$, $D=C(2\lambda -1)$, $%
S=C(\lambda -1)(3\lambda -1)$. In this paper the used material is aluminum.
The corresponding parameters are $\rho _{0}=2700$ kg/m$^{3}$, $E=69$ Mpa, $%
\nu =0.33$, $\sigma _{Y0}=120$ Mpa, $E_{\tan }=384$ MPa, $c_{0}=5.35$ km/s, $%
\lambda=1.34$, and $\gamma_0=1.96$ when the pressure is below $270$
GPa \cite{DBM}.

Before the numerical experiments, we validated our code by several
benchmark simulations. The first one involves a cylinder rolling on
an inclined rigid plane. The second involves the collision of two
elastic spheres. The third involves a copper Taylor bar impacting to
a rigid wall. The fourth is to simulate the process of the collision
between four identical spheres. All the simulation results agree
well with the analytic or experimental ones\cite{PXFGF2007}.

In our numerical experiments the shock wave to the target material
is loaded via either of two equivalent means, impacting with another
body in the vertical direction or by a rigid wall at the bottom. The
horizontal boundaries are treated as rigid walls. The initial
configuration of material particles are set to be symmetric about
the central vertical line. Such a configuration corresponds also to
a real system composed of many of the simulated ones aligned
periodically in the horizontal direction. In this paper we focus on
the two-dimensional simulations.

\section{Results of numerical experiments}

As the first step, we set a single cavity in the metal material. Due
to the boundary conditions mentioned above, such a simulation model
corresponds also to a very wide system with a row of cavities in it.
We study various cases where the shock wave changes from strong to
weak. In the cases with strong shock, we concern the jet creation
and the distribution of the ``hot spots''. When the cavity is close
to the free surface, we check whether or not there are material
ejected out of the free surface. In the cases with weak shock, we
study the effects of cavity size, distance from the cavity center to
the impacting face, the initial yield stress of the material, etc,
to the collapsing procedure.

\begin{figure}[tbp]
\centering \caption{(See Fig1.jpg))Snapshots of collapse of a single
cavity under strong shocks. From black to white the gray level in
the figure shows the increase of local temperature denoted by the
plastic work during the deformation procedure. The spatial unit is
$\mu m$. The unit of work is mJ. (a)t=2 ns, (b) t=5 ns. }
\end{figure}

\begin{figure}[tbp]
\centering \caption{(See Fig2.jpg)Snapshots of collapse of a single
cavity under a strong shock. For the pressure contour, from black to
white the value increases. The corresponding times at Figs.(a)-(f)
are 1.2, 1.6, 1.8, 2.0, 2.2, 12 ns, respectively. The spatial unit
is mm. The unit of pressure is Mpa.}
 \caption{(See Fig3.jpg)Configurations with
local temperature denoted by the plastic work during the
deformation. The unit of work is mJ. Figs.(a)-(f) here correspond to
Figs.(a)-(f) in Fig.2, respectively. }
\includegraphics*[scale=0.8]{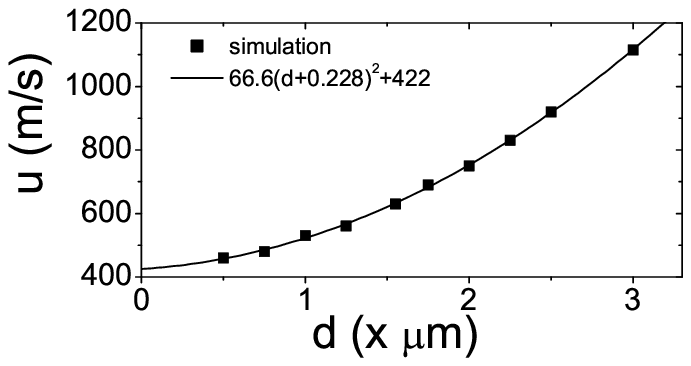}
\caption{Critical impact velocity versus the thickness of the downstream
wall of the cavity. The symbols are simulation results and the solid line is
the fitting result.}
\end{figure}


Figure 1 shows a snapshot for a case with very high impacting speed. The
width of the simulated system is 10 $\mu m$. Two metal bodies with symmetric
configurations collide. The speed of each is 1500 m/s. That is to say, the
relative impacting speed is 3000 m/s. Thus, the initial pressure to one body
loaded by the shock is about 30 GPa being less than the critical value, 270
GPa. The global procedure can be described as follows: (i) When the shock
wave arrives at the upper-stream wall of the cavity, plastic deformation
begins to occur. The shock waves at the two sides of the cavity move
forwards to the free surface. The shock speed at the two sides is larger
than the deforming speed of the upper-stream wall of the cavity. (ii) The
upper-stream wall continue its collapse, a configuration with a turned ``C"
appears. The material projected into the cavity makes a jet.
\textquotedblleft Hot spot\textquotedblright occurs at the tip region of the
jet. (See Fig.1(a).) (iii) The speed of jet increases with time. The tip of
the jet catches up, then exceeds the shock waves at its two sides. (iv) The
jetted material impacts the down-stream wall of the cavity, results in a
pair of vortices rolling in opposite directions. The ``hot spots" appear at
the centers of vortices. (See Fig. 1(b).)

In the case where the cavity locates near the upper free surface, if
the shock wave is strong enough, the material projected into the
cavity will become a jet and hit the downstream wall and break it.
This procedure makes another dynamical picture of ejection by shock.
Such a behavior has been observed in experiments and has some
potential applications. We show such a dynamical procedure in Fig.2,
where the shock wave is loaded by colliding with rigid bottom wall.
The simulated system here is 10 $\mu m$ in width. The initial radius
of the cavity is 1.5 $\mu m$. The distance between the cavity center
to the upper free surface is 4.5 $\mu m$. The impacting speed of the
metal body to the rigid wall is 1120 m/s. The corresponding times in
Figs. (a)-(f) are 1.2 ns, 1.4 ns, 2.0 ns, 2.4 ns, 4.4 ns, and 12.0
ns, respectively. Fig.2(a) shows a snapshot where the shock wave has
passed most part of the cavity. The upper-stream part of cavity has
been substantially deformed and some material have been projected
into the cavity. Up to the time $t=$1.6 ns, as shown in Fig.2(b),
the cavity has been nearly filled with the jetted material; the
compressive wave is arriving at the free surface from the two sides
and rarefactive waves will be reflected back. The reflected
rarefactive waves decrease the pressure in the influenced region and
cavitation occurs around the region of the original cavity and at
the two sides of the jet path. (See Figs. 2(c)-2(d).) Compared with
those at the two sides, material in the middle is in a much higher
pressure and has much more pronounced kinetic energy so that it is
significantly distorted. The newly created cavities coalescence and
become a larger one with time. (See Fig.2(e).) If the upper wall of
the newly created cavity possesses enough kinetic energy, it will be
be broken. (See Fig.2(f).) Fig. 3 shows the corresponding
configurations with temperature. In Fig.(a) the temperature of the
compressed region around the cavity is higher that in other region.
The ``hot spot'' is at the tip of the metal tongue. In Figs. (b)-(c)
the ``hot spot'' occurs at regime hit by the metal tongue. After the
appearance of the new cavity, ``hot spots'' locate at the inner wall
of the cavity, especially the upper and bottom walls. (See Figs.
(d)-(f).) If there are metal material ejected from the upper free
surface depends on the initial impacting speed $u$ and the width of
the upper wall of the original cavity $d$. The critical impacting
speed $u$ increases parabolically with $d$. Fig. 4 shows the
simulation results and fitting curve.

\begin{figure}[tbp]
\centering \caption{(See Fig5.jpg)Snapshots of collapse of a single
cavity under a weak shock. From black to white the gray level in the
figure shows the increase of local pressure. The spatial unit is mm.
The unit of pressure is Mpa. (a) t=1.0 ns, (b) t=1.6 ns, (c) t=2.2
ns, (d) t=3.0 ns, (e) t=5.4 ns, (f) t=16.0 ns.} \caption{(See
Fig6.jpg)Configurations with local temperature denoted by the
plastic work during the deformation. The unit of work is mJ. }
\end{figure}

\begin{figure}[tbp]
\centering \caption{(See Fig7.jpg)Transition of symmetry of
collapsing. (a) Asymmetric in vertical direction near the impacting
face; (b) Nearly symmetric collapse. The gray level in the figure
corresponds to the plastic work. The spatial unit in the figure is
mm. The unit of energy is mJ.}
\end{figure}

\begin{figure}[tbp]
\centering
\includegraphics*[scale=1.4]{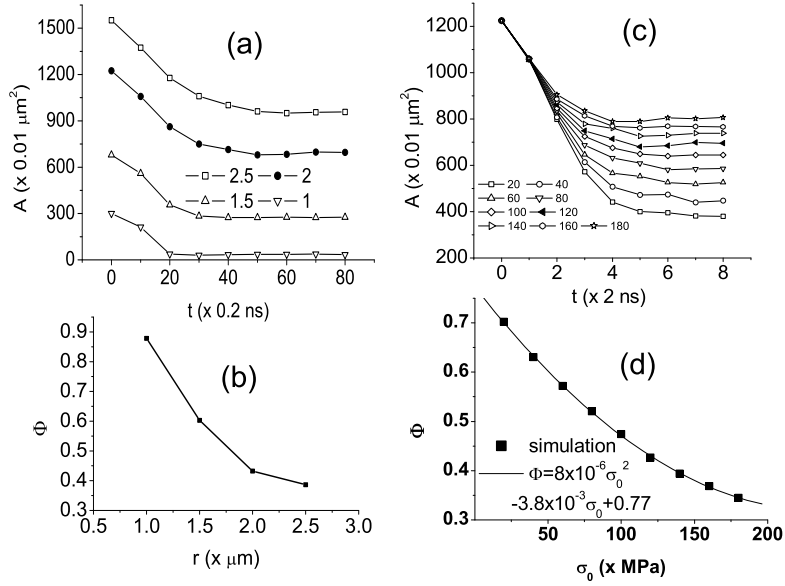}
\caption{Effects of cavity size [(a) and (b)] and initial yield [(c)
and (d)] to the collapsing procedure. (a) Area of cavity versus time
for different initial radius. What shown in the legend are the
initial radii, the unit is $\mu$ m; (b) Collapsibility versus
initial radius of cavity; (c) Area of cavity versus time for
different initial yield stresses. What shown in the legend are the
initial yield stresses, the unit is Mpa. (d) Collapsibility versus
initial yield stress.}
\end{figure}

With the decrease of impact speed, the collapse becomes slower. When
the impact speed decreases to about 200 m/s, the dynamical picture
has been significantly different: The cavity can not be collapsed
completely, and the final configuration around the cavity varies
from (nearly) symmetric to asymmetric if we change the distances
from the cavity center to the impacting surface. In Fig. 5 we show
an ``abnormal" asymmetric dynamical picture where the cavity
collapsed less in the initial shock direction. Here the width of the
simulated system is 10 $\mu m$. The initial radius of the cavity is
1.5 $\mu m$. The distance between the cavity center to the impact
surface of the metal body is 1.9 $\mu m$. To understand the observed
behavior, we ``recover" or ``magnify" the system in such a way: The
rigid walls at the two sides and at the bottom of the computational
region can be regarded as ``mirrors". In other words, the system is
symmetric about the ``mirrors". The distance between two successive
cavities in the horizontal direction is $d_H =10 \mu m$; while the
distance between the cavity in the computed body and the  fictitious
one symmetric about the impacting face is $d_V = 3.8 \mu m$. The
cavities reflect rarefactive waves. Therefor, the pressure in
between decreases. Since the distance $d_V$ in vertical direction is
much less than $d_H$ in horizontal direction, the rarefactive waves
in between the two cavities in vertical direction reflected more
frequently. Correspondingly, the pressure in this region is much
lower than those in the surrounding region. This results in the less
collapsing of the lower part of the cavity. When the shock waves
arrive at the upper free surface, rarefactive waves are reflected
back towards the cavity. This is a second reason for the cavity to
collapse less in the vertical direction. Compared with those from
the fictitious cavity below the bottom, the reflected rarefactive
waves from the upper free surface is much wider. This explains why
the collapsing of the lower part of the cavity is more pronounced.
Fig.6 shows the corresponding configurations with temperature. The
hottest region appears in the region below the cavity. In Figs. 5
and 6, Figs. (a)-(f) correspond to 1.0 ns, 1.62 ns, 2.2 ns, 3.0 ns,
5.4 ns and 16 ns, respectively. From Figs. 5(a) and 6(a) it is clear
that, around this time, although the rarefactive wave decreased the
pressure within the influenced region, but the temperature there is
higher than those in other parts of the system. The reason is that
the waves there did more plastic work. Since the cavity here
corresponds to vacuum, pure fluidic models are not able to capture
such a phenomenon.

The temperature in the ``hot spot" increases in the course of
collapsing. If we further decrease the distance between the cavity
and the lower impacting face, the lower part of the cavity will be
collapsed more pronouncedly. Fig. 7(a) shows the final steady state
for a case where the lower boundary of the cavity just locates at
the impacting face. In contrast, if we increase the distance, the
collapsed cavity will be more symmetric. Fig. 7(b) shows a case
where the collapsing is nearly isotropic.

We studied cases with various sizes of cavity. It is found, for the
investigated cases, that the collapsibility becomes larger when the
initial size of the cavity is reduced. The variation of area of the
cavity with time is shown in Fig. 8(a), where four kinds of radius
are used. They are 2.5 $\mu m$, 2.0 $\mu m$, 1.5 $\mu m$, and 1 $\mu
m$, respectively. If we define the collapsibility as
$\Phi=(A_0-A)/A_0$, then it is clear that $\Phi$ decreases as the
cavity becomes larger, where $A_0$ and $A$ are the areas of the
cavity in the initial and final states. (See Fig. 8(b).) In order to
study separately the effect of individual material characteristics
to the collapsing procedure, we changed the initial yield stress of
the material under the condition that all other parameters are
fixed. The corresponding collapsing procedures are shown in
Fig.8(c), where the numbers in the legend indicates the used initial
yield stresses. The unit is MPa. Fig. 8(d) shows that the
collapsibility decreases parabolically if the initial yield becomes
larger.

\section{Conclusion}

The collapse of cavities under shock is investigated by the
material-point method. What mainly concerned in the present paper
include behaviors ignored by pure fluidic models and those quick
distortion procedures which are generally difficult for the
traditional finite-element method \cite{FE}. We studied the
relations between symmetry of collapsing and the strength of shock,
other coexisting interfaces, as well as hydrodynamic and
thermal-dynamic behaviors. In the case with strong shock, the
procedure of jet creation in the cavity in studied. The jetted
material hits the downstream wall and results in two vortices
rolling in opposite directions. The ``hot spots" occur in the two
centers of vortices. When the shock is strong enough, the jet
material with high kinetic energy hits and breaks the downstream
wall. The critical impact speed for such a phenomenon increases
parabolically with the thickness of the downstream wall. In the case
with weak shock, it is found that the cavity can not be collapsed
completely and the cavity may collapse in a nearly isotropic way.
The transition of symmetry in the course of collapsing is relevant
to the size of the cavity, the distances to the walls, as well as
the impacting speed, etc. The distribution of ``hot spots" in the
shocked material changes for different collapsing procedures. Since
we use the Mie-Gr\"{u}neisen equation of state and the effects of
strain rate are not taken into account, the behavior is the same if
the spatial and temporal scales are magnified in the same way. The
results obtained in the paper will help to understand better the
course of ignition of inhomogeneous explosives, fatigue and erosion
of metal materials, etc.


\ack{ We warmly thank G. Bardenhagen, Haifeng Liu, Shigang Chen,
Song Jiang, Xingping Liu, Xijun Yu, Zhijun Shen, Yangjun Ying, Guoxi
Ni, and Yun Xu for helpful discussions. We acknowledge support by
Science Foundation of Laboratory of Computational Physics, IAPCM,
National Science Foundation (No.10472052) and National Basic
Research Program (No. 2002BC412701) of China.}

\end{document}